# Formulation Graphs for Mapping Structure-Composition of Battery Electrolytes to Device Performance

Vidushi Sharma[*], Maxwell Giammona, Dmitry Zubarev, Andy Tek, Khanh Nugyuen, Linda Sundberg, Daniele Congiu, Young-Hye La[**]

*IBM Almaden Research Center, 650 Harry Rd, San Jose, CA, USA*

*Email: vidushis@ibm.com, **Email: yna@us.ibm.com

**Abstract**

Advanced computational methods are being actively sought to address the challenges associated with the discovery and development of new combinatorial materials such as formulations. A widely adopted approach involves domain-informed high-throughput screening of individual components that can be combined into a formulation. This manages to accelerate the discovery of new compounds for a target application but still leaves the process of identifying the right 'formulation' from the shortlisted chemical space largely a laboratory experiment-driven process. We report a deep learning model, Formulation Graph Convolution Network (F-GCN), that can map the structure-composition relationship of the formulation constituents to the property of liquid formulation as a whole. Multiple GCNs are assembled in parallel that featurize formulation constituents domain-intuitively on the fly. The resulting molecular descriptors are scaled based on the respective constituent's molar percentage in the formulation, followed by integration into a combined formulation descriptor that represents the complete formulation to an external learning architecture. The use-case of the proposed formulation learning model is demonstrated for battery electrolytes by training and testing it on two exemplary datasets representing electrolyte formulations vs. battery performance - one dataset is sourced from literature about Li/Cu half-cells, while the other is obtained by lab experiments related to lithium-iodide full-cell chemistry. The model is shown to predict performance metrics like Coulombic Efficiency (CE) and specific capacity of new electrolyte formulations with the lowest reported errors. The best-performing F-GCN model uses molecular descriptors derived from molecular graphs that are informed with HOMO-LUMO and electric moment properties of the molecules using a knowledge transfer technique.

## 1. Introduction

Machine Learning (ML) methods have transformed the way materials scientists approach and survey material design spaces. Driven by the rise of high-throughput electronic structure calculation-derived datasets [1-3], ML methods now present fast and accurate prediction of physio-chemical properties of molecules [4-6], solid-state materials [7,8] and combinatorial spaces such as interfaces [9,10]. In recent years, generative artificial intelligence (AI) has further taken a leap at hypothesizing novel molecules with well-defined functionalities for different applications [11-13]. However, the successful application of any discovered novel molecule at the device level is not guaranteed as the performance of a device goes beyond the sum properties of individual components and mostly relies on the interplay of complex interactions between all its constituent components. There exists a gap between the discovery of new molecules with generative AI and their actual application, which can be bridged with a machine learning or deep learning method representing mixed material systems (Figure 1).

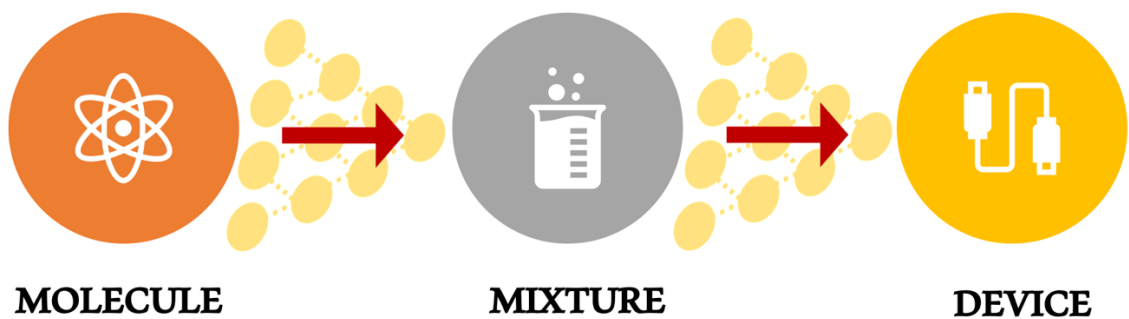

**Figure 1:** Learning models for mixed material systems, such as formulations, bridge the gap between the discovery of new molecules and their actual application in the device.

Liquid formulations are one example of such mixed material systems that are a big part of many industrial sectors like pharmaceuticals, automotive materials, food science, coatings, and personal care products [14-16]. So far, high-throughput screening has been successful in accelerating the search for new individual compounds in multi-constituent systems [17, 18]. Yet, our current methods have fallen short of directing the complete design of new materials formulations. 'Data' is a major pre-requisite before scientists can leverage data-driven methods for formulation space. Despite advancements in material simulation techniques, it remains computationally demanding to simulate the properties of whole liquid formulations. The viscosity of mixtures, salt solubilities in mixed solvents, and density of liquid mixtures are a few popular open problems that remain challenging to solve with simulations [19-23]. Therefore, experimental data continue to be a major fuel to ML engines used in formulation-based predictions [24, 25]. Gathering high-quality formulation data from experimental procedures to train ML models is a laborious process, especially when the problem statements cross over to device-level applications such as in the case of battery electrolytes. The discovery and optimization of battery electrolytes has become the subject of paramount importance to climate and sustainable technologies, for it will accelerate decarbonization across all economic sectors. Liquid electrolytes in modern energy storage devices typically involve one or more organic solvents and one or more salt additives [26]. The formulation of constituent salt-solvents in an electrolyte has been shown to have significant impacts across many cell performance outcomes such as capacity retention, rate performance, and cycle life [27-29]. A straightforward approach to predict the performance metrics of formulated products such as battery electrolytes is by supervised learning models like regression [30-32] that take selective features of constituent materials as input [33, 34]. These input features can be domain intuitive properties of electrolyte constituents like redox potentials, dielectric constants, ionic conductivity, solubility, and viscosity. The process requires an intelligent input feature selection process and reliable methods of calculating them [35], thus bringing forth the critical issue of error transmission from methods estimating the input features to the outcomes of the ML model predicting formulation properties. In the latest attempt by Kim et al. [33], a list of elemental composition (primarily oxygen and fluorine) features is intelligently defined to predict the Coulombic efficiency (CE) of lithium metal battery electrolytes using regression models trained on empirical Li/Cu half-cell data. Though the study is an excellent example of a data-driven electrolyte design strategy, the method may not be generalizable to other formulation problems as it skims over crucial structural and compositional information about formulation constituents.

In the present work, we propose a formulation graph convolution network (F-GCN) model to predict properties of the formulated products, such as battery electrolytes, based solely on constituent structures and compositions. Graphs present a natural framework for characterizing the attributes of materials. They offer feature engineering on the fly by learning from the structure, hence bypassing additional feature engineering steps, streamlining uncertainty propagation, and accounting directly for the geometric and structural foundations of the interactions in molecular systems. Molecular systems are represented as graphs to a learning algorithm of graph convolution networks (GCN) [36], that falls under the umbrella of graph neural networks (GNNs) [37]. GCNs are proven to be effective for molecular level predictions and are used to map molecular fingerprints [36, 38], study targeted interactions between two compounds [39, 40], predict molecular properties [41-44], learn chemical reactions networks [45], and map atom networks in the crystals [46, 47]. Graph models can be further extended to a global scape by either incorporating dense layers or by structuring an external graph called a dual graph. Dual graphs effectively map chemical interactions in dense chemical spaces such as crystalline materials [45, 48], and therefore remain suited for dense solid-state systems. The framework of the proposed F-GCN model maps the structure of the formulants to the outcome metrics based on their respective molar compositions. The framework simplifies the process of input feature selection by using single-line molecular identifiers (SMILES) along with molar compositions as input, reduces the requirements for a large formulation dataset by incorporating a two-stage learning process, opens the possibility of imparting domain knowledge to a deep learning model for enabling application based customization, and most importantly, defines the concept of formulation descriptor, i.e. a vector representation of liquid formulations to machine learning models. We use the Li/Cu half-cell dataset from the study by Kim et al.[33] containing a total of 160 electrolytes to benchmark the proposed methodology. Next, the model is demonstrated to predict the performance of electrolytes for a more intricate system of lithium-iodide (Li-I) full battery, post-training on variable electrolyte formulation data from the same system.

## 2. Formulation Graph Convolution Networks (F-GCN)

To map the performance of a formulated material such as a battery electrolyte, it is essential to take the composition of the constituent materials into account. Figure 2 illustrates the architecture of the F-GCN model used in this work to map the structure and composition of electrolyte formulants to the measured performance outcomes of the devices that use those formulations. The model is comprised of two learning architectures: graph convolution networks (GCN) and deep neural networks (DNN). The model takes SMILES (Simplified Molecular Input Line Entry System) [49] of the electrolyte components as input along with their molar composition as a percentage of the total electrolyte formulation (denoted as Molar % in Figure 2). Formulant SMILES are converted to molecular conformations using an RDKit package [50] and are then transformed into molecular graphs for GCNs. A molecular graph contains $N$ nodes represented by $f_a$ features, where $N$ is the number of atoms in the molecule and $f_a$ is a numeric vector of size 100. A binary one-hot encoded matrix is used as $f_a$ for each node, where a non-zero entity (atom's electronegativity in the present work) is located at the $n^{th}$ position, with $n$ being the atomic number of the atom (see Supporting Information SI-1). The vector size of $f_a$ is set to be 100 assuming that electrolyte formulants will not contain atoms with atomic numbers larger than 100. Symmetric adjacency matrix $A$ designates the connections between the nodes within the graph including the bond types [43]. GCNs [51] are analogous to convolution neural networks (CNN) used for images [52] where each layer convolutes the previous layer's features and produces a new set of features combining the information from the neighboring pixels. GCN extends this concept to graphs by performing the convolution functions on each node until the neighborhood of the entire graph is represented in the feature set. GCNs have previously been used as quantitative structure-activity relationship (QSAR) models due to their ability to extract trainable features from unstructured graphs without any prior feature engineering step [53-56]. F-GCN uses GCN for feature extraction from formulant structures into a molecular descriptor shown in red in Figure 2(c). Convolution actions are performed on each molecular graph by hidden convolution layers as per Kilf and Welling's theory [51], to modify features of nodes ($f_a$). During convolution, the model learns the chemical neighborhood of each atom and updates the respective nodes. As the number of convolutions increases, the chemical neighborhood for each atom is expanded. Post convolutions, each atom is represented by a modified $f_a$ which is informative of the chemical neighborhoods within the molecule. The modified $f_a$ of all the atoms comprising a molecule are averaged to derive a graph representation (*GR*) as a 1-D descriptor (vector size = 100) for the whole molecule. This ultimate layer of the GCN model can be assumed to encode important chemical characteristics of the molecule and be used as a molecular descriptor input for external learning architecture [57].

The distinctive feature of the F-GCN model relative to prior models [58-60] is the ability to combine molecular descriptors into a formulation descriptor based on the composition (illustrated in Figure 2(c-e)). Considering molecular descriptor *(GR)* to be a saturated representation (100%) of an individual molecule, we can alter molecular representations in the model by scaling *GR* with each molecule's molar percentage in the formulation as per Equation *(1)*:

$$GR'_{mol1} = \omega \cdot GR_{mol1} \qquad (1)$$

where $GR_{mol1}$ is the $GR$ matrix of molecule 1, $\omega$ is the fraction of the molecule's molar percentage in the formulation ($\omega$ is 0.50 for 50%) and $GR'_{mol1}$ is the scaled $GR$ of molecule 1. The scaled GR ($GR'_{mol}$) of the molecular constituents are then summed together to create a unified descriptor for the formulation ($DS_{for}$) as shown in Equation *(2)*:

$$DS_{for} = \sum_{i=1}^{j} GR'_{mol_i} \qquad (2)$$

Here, *j* is the number of total formulants in the formulation. $DS_{for}$ can be assumed to contain cumulative features of all formulants. $DS_{for}$ is input to an external learning architecture that maps to the final formulation property, i.e. battery performance metrics here. We use dense feed-forward neural networks (DNN) for external learning architecture, to map complex non-linear relationships in the formulation space. One may choose a simplified learning algorithm such as linear regression, kernel methods, or random forest regressor to map the formulation descriptor ($DS_{for}$) to the output label, based on the intricacy of the problem.

A Python version of the F-GCN model is developed using the *Keras* API [61] with TensorFlow [62]. The model has a customizable count of six parallel GCNs for complex electrolyte formulations that comprise at most 6 molecular components. Each GCN takes a molecular graph and performs a set of four convolutions before averaging the modified $f_a$ into $GR$. A non-linear activation function *tanH* is applied to the output from each convolution layer before passing the new $f_a$ to the next layer. GCN architecture stays consistent in the F-GCN model post-optimization (see Supporting Information SI-2 for complete details of GCN). Meanwhile, DNN hyperparameters are custom-tuned for different performance metrics. Considering the high degrees of freedom in the model and the small formulation dataset derived from experiments, F-GCN is trained with batch size 1 until convergence is reached. The model is converged during training with Adam optimizer [63] with the learning rate as small as 0.0001. A detailed description of the F-GCN framework in TensorFlow is shared in Supporting Information SI-3.

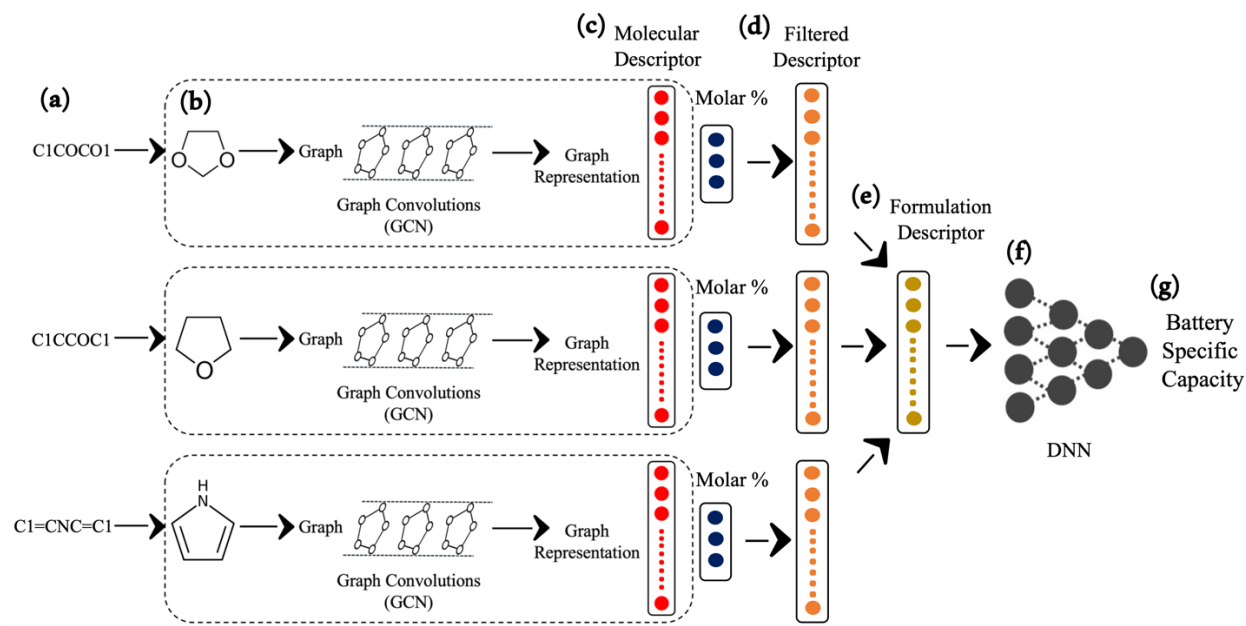

**Figure 2:** Description of Formulation Graph Convolution Network (F-GCN) model for 3 component formulations: **(a)** SMILE strings of constituent molecules as input to F-GCN, **(b)** Conversion of inputs to molecular graphs for Graph Convolution Network (GCN), **(c)** Graph representations (*GR*) as output from GCNs, also referred to as Molecular Descriptors, **(d)** Molecular descriptors that are scaled based on the respective molar percentage in the formulation, **(e)** Formulation descriptor (***DS_{for}***) representing features of the complete formulation to external learning network, **(f)** DNN that is the external learning architecture training on formulation data, **(g)** Formulation property or label of interest i.e. Battery Performance for electrolytes.

## 3. Electrolyte Datasets

### *3.1. Li/Cu Half-cell Dataset from Literature*

The F-GCN model is trained and benchmarked with a dataset of Li/Cu half-cell-based electrolyte formulations and their respective coulombic efficiencies (CE) obtained from the reference [33]. Kim et. al [33] curated the dataset from the literature and transformed CE to the logarithmic format (LCE) such that it numerically amplifies the change in the output with respect to the changes in the electrolytes. CE is the ratio of discharge and charge capacity of the battery and is an important metric of battery performance. Battery can suffer from CE loss over time due to electrolyte and electrode decomposition. Therefore, electrolyte engineering has become a major strategy to improve CE.

There are 147 electrolyte formulations in the acquired training dataset and an additional 13 electrolyte formulations, that are used to evaluate the performance of the model as test data (tabulated in Supporting Information SI-4). Trainable 147 electrolyte formulations in the dataset consist of 2 to 6 electrolyte components in each, described by SMILES. To structurally represent the maximum electrolyte components in the model, a set of 6 GCNs is applied in parallel. In electrolyte data points containing less than 6 formulants, the remaining formulants are defaulted to be water, and their molar composition is set to 0 (dummy featurization). This ensures that the model can handle data points with variability in the formulant count. Figure 3 classifies the training dataset based on the number of electrolyte components and presents the distribution of LCE output in the dataset with the box plots. In Figure 3(a), it is evident that a major fraction of the dataset has 3-4 formulants, while only a small count of formulations in the dataset (2/147) contains all 6 components. This implies that some of the graphs (GCN) in the model learn dummy molecules for most training unless the dataset is largely augmented. For a reliable representation of formulation space, F-GCN must robustly predict the output label despite the translation in the sequence of formulants in the dataset, i.e. solvents and salts. This could be achieved by the strategy of knowledge transfer using easily computable simulation data as discussed in the next section.

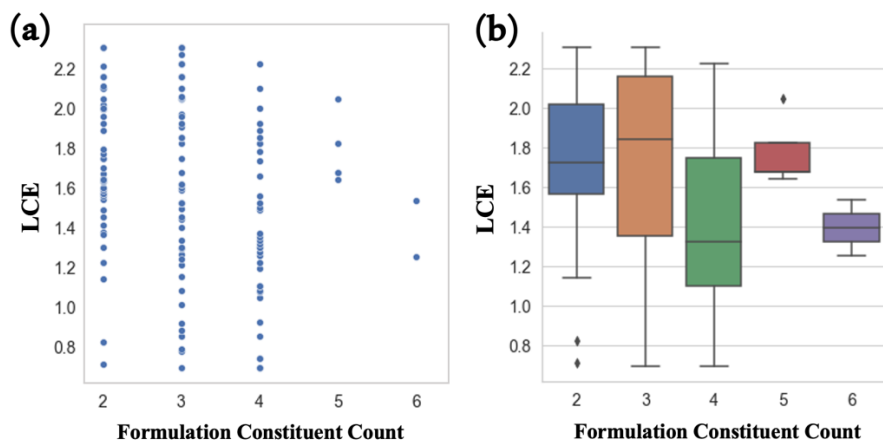

**Figure 3:** Analysis of training data from Li/Cu half-cells. **(a)** Distribution of electrolyte formulation data points based on the formulation constituent count and LCE values. **(b)** Box plots showing the distribution of LCE outputs for training data based on formulation constituent count. The outer whiskers represent the minimum and maximum values, the central line represents the median, and the colored box represents the 25$^{th}$ to 75$^{th}$ percentile of the data. The data points outside the outer whiskers are outliers observed in the data.

*3.2. Simulation Dataset*

Knowledge Transfer (also called Transfer Learning (TL)) is a general approach in deep learning where a model trained for a specific task is re-used for another task. TL is deployed in deep learning models to overcome conditions of lacking datasets, necessities to embed domain knowledge, or to bring more transferability for broad application. Knowledge transfer is an additional step in the F-GCN workflow to overcome the limitations of a small experimental dataset. With higher numbers of formulants (GCN = 6), dimensionality and complexity of the F-GCN model increase, therefore finding the dataset of 100-150 formulations insufficient in capturing the necessary physical, chemical, and synergistic influences in the formulation towards performance label. To overcome the limitations of scarce data and bring more transferability to the F-GCN model, we pre-train GCN on a wider dataset of molecules to enable improved accuracies of F-GCN in capturing general features of chemical compounds. There exists chemical and physical data for over 1 billion molecules in open-sourced databases such as ZINC [64], PubChem [65], and QM9[66] which can be successfully used for pre-training of structural models such as GCN [67-69]. However, training models with this large pool of available open-sourced molecular data requires extensive computational time and a separate focused development. Herein, we demonstrate the application of the concept by pre-training GCN with a smaller in-house created quantum chemical dataset of 500 molecules that consists of commonly found solvents and salt additives in Li metal batteries. Density Functional Theory (DFT) calculations are performed with GAMESS software using functional B3LYP and basis set 6-311 G (d,p) for all 500 molecules to calculate their frontier orbitals (HOMO-LUMO levels) and electric dipole moments (EM). The molecules and their simulated results are provided along with the Supporting Information. The distribution of simulation data used for transfer learning is shown in Figure 4. Box plots in Figure 4 indicate the presence of some outliers in the LUMO and EM data. It is crucial to indicate that these outliers were not excluded from the training. For the knowledge transfer, GCN is trained for the molecule's physio-chemical property labels (HOMO-LUMO energy levels and electric moment) as illustrated in Figure 5. GCN architecture used for pre-training is the same as described to be part of F-GCN in Section 2. In the process of pre-training, GCN learns the effects of local chemical neighborhoods in the molecule towards the label. Thereby, molecular descriptors are created that encode molecular information of the targeted specific property (HOMO-LUMO energy levels and electric moment here) through the process of back-propagation [57]. These informed molecular descriptors are used to represent formulants in the F-GCN framework, to be subsequently combined into a formulation descriptor.

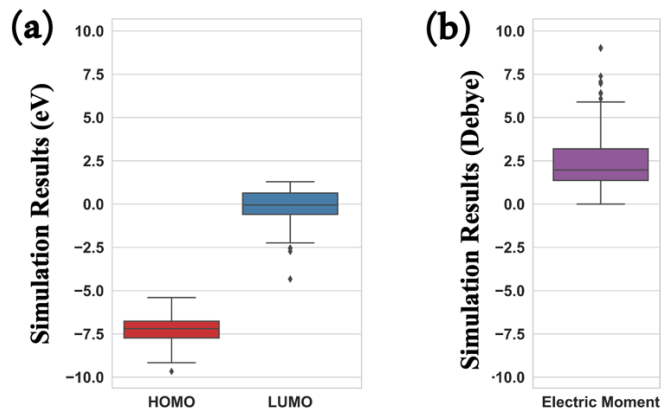

**Figure 4:** Box plots showing the distribution of simulation outputs used for creating informed molecular descriptors using knowledge transfer. **(a)** Distribution of HOMO-LUMO energy levels for 500 salts and solvents. **(b)** Distribution of electric moment of 500 salts and solvents. The simulation data is used for pre-training GCN.

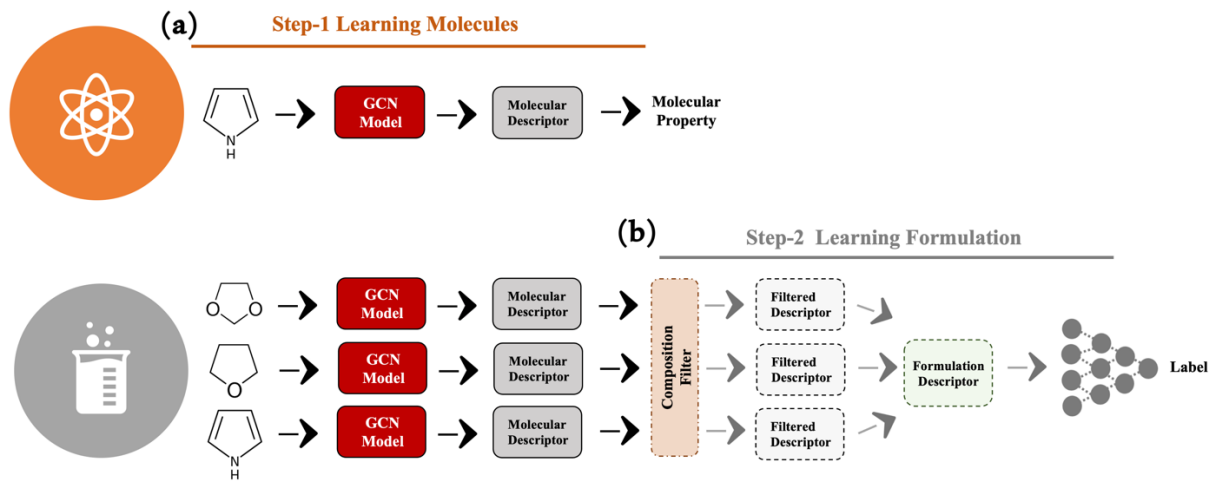

**Figure 5:** Knowledge transfer step to create informed molecular descriptors from molecular structures in F-GCN workflow for electrolyte formulations: **(a)** GCN is trained on output labels HOMO-LUMO (HL) energy levels and electric moment (EM). **(b)** Pre-trained GCN is used in the F-GCN network, with GCN weights set to non-trainable. GCN outputs uniformly represent formulants in the F-GCN model to external learning architecture that maps them to the overall battery performance.

*3.3. Li-I Full-cell Battery Dataset*

A dataset of a total of 125 electrolyte formulations and specific capacities is obtained experimentally for Li-I battery coin cells [70] represented in Figure 6(a) by performing cycling tests at 1 mA/cm$^2$. While the theoretical capacity of the battery (211 mAh/g-cathode for present Li-I battery [70]) is determined by the redox couple of cathode and anode materials, electrolyte plays a key role in maximizing the battery's practical capacity, along with other performance metrics such as charge rate and cycle life. Especially for lithium-metal batteries based on conversion chemistries, such as a Li-I battery, electrolytes should be formulated carefully to prevent the shuttling of active species and deleterious electrolyte decomposition reactions. Out of 125, 111 electrolyte formulations are used for the training of F-GCN while random 14 electrolyte formulations are retained as test datasets that are used to validate the performance of a trained F-GCN model. The box plots in Figure 6(b) show the distribution of capacity labels obtained from experiments for the training and test datasets. Complete details of the data collection procedure, cell assembly, and experimental set-up are provided in the Supporting Information SI-5 and SI-6. As seen in Figure 6(b), the training data is heavy with electrolyte formulations having specific capacities high (above 80 mAh/g) and poor (~0 mAh/g), yet the average specific capacity of the electrolyte formulations noted in training data is 59 mAh/g. On the other hand, the average capacity noted for the test dataset (14 electrolyte formulations) is 72 mAh/g. Note that all the cells had a small degree of variability in cathode formulations in terms of carbon additive, binder, and active material loading percentage (45-55 wt%). For standard deviation in the experimental capacity as the result of cell-to-cell variability in Li-I full battery coin cell, see Supporting Information SI-7.

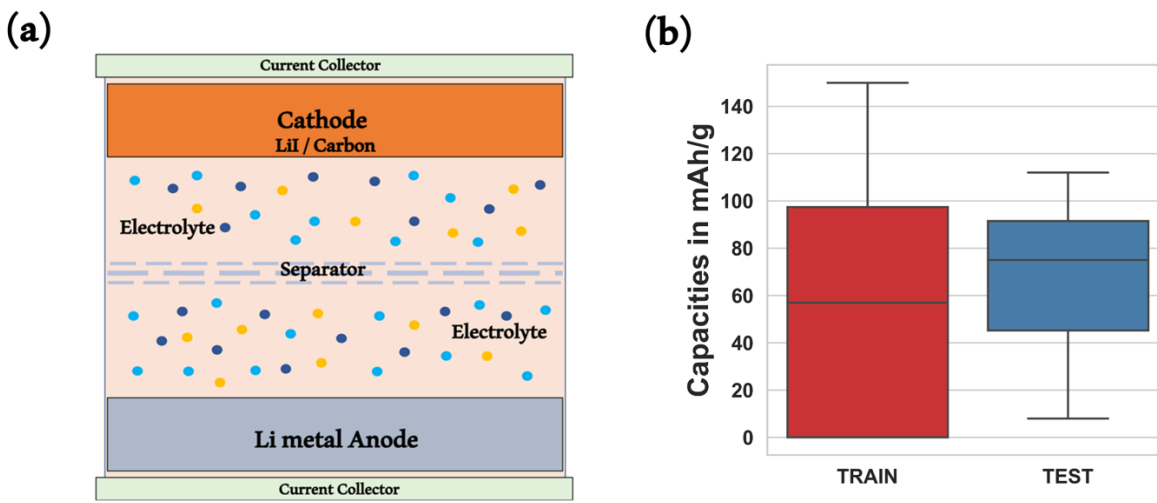

**Figure 6: (a)** Schematic representation of Li-I battery cell that is used to collect electrolyte formulation vs. specific capacity experimental dataset in the laboratory. **(b)** The box plots depict the distribution of output labels, i.e. specific capacity obtained from experiments that are used for training and evaluating the performance of the F-GCN model. The outer whiskers represent the minimum and maximum values, the central line represents the median, and colored box represents 25th to 75th percentile of the data.

## 4. Results and Discussion

The goal of the F-GCN model is to featurize a formulation based on concerted representations of molecular constituents and compositions in a mixture. To accomplish this with learning models that have thousands of trainable parameters, a variety of training samples are required. However, obtaining such variant training data is often not feasible, especially when relying on complex empirical procedures involving manual fabrication and evaluation of individual devices. As a result, models can fail to learn generic concepts in the input space. To accomplish meaningful learning with a small formulation dataset, F- GCN is trained within two domains: learning the chemical underpinnings of molecules and learning compositional correlations of molecules in formulations with the final performance. To demonstrate the effectiveness of this segmented learning, we train two separate F-GCN models with each of the two electrolyte datasets: F-GCN with no Knowledge Transfer (no-TL F-GCN) and F-GCN with pre-trained graphs (TL F-GCN). The results from both the models are presented in this section.

*4.1. LCE Prediction*

We train two F-GCN models (no-TL F-GCN and TL F-GCN) utilizing a dataset of electrolyte formulations vs. LCE, which was sourced from prior literature [33]. A version of the F-GCN model is optimized for predicting LCE by tuning hyperparameters such as number of hidden layers in DNN, node count, and activation function by a trial-and-error approach. The performance of different hyperparameters considered for evaluation is further detailed in Supporting Information SI-8. For hyperparameter tuning, 20 % of the dataset is split for validation and testing. The final external learning model (DNN) is a 3-layered perceptron with each layer having 25-10-1 nodes, the last of which is an output label. Robust convergence during the model training is noted when the rectified linear activation unit (*relu)* function is applied to the hidden layers of DNN while the last layer is connected to the output linearly. It took approximately 5,000 epochs for the model to reach convergence. The performance of the model is evaluated by calculating the mean squared error (MSE) in the predicted outputs (LCE).

Figure 7 summarizes the overall performance of the two F-GCN models (no-TL F-GCN and TL F-GCN). The trained models are tested on 13 test electrolyte formulations and their respective experiment-derived results as reported by Kim et al [33]. Figure 7(a) compares loss (MSE) calculated for training and validation data during the learning process by the F-GCN model (with the best-developed network and no transfer learning) for 5000 epochs. Curves for both the loss functions are sloping towards zero with training and validation MSE being 0.0038 and 0.1704, respectively. This demonstrates a good learning curve for the model. The trained model is further used to predict LCE for unseen 13 electrolyte formulations present in the test dataset. The parity plot in Figure 7(b) indicates the correlation between the LCE values predicted by the no-TL F-GCN model and the experimental results. The straight red line in the plots maps the benchmarked experimental values and scatter points denote the predicted LCE values. The difference of the scatter points from the straight line indicates the errors in the predicted LCE. The MSE value for the plot is indicated in the figure to be 0.61.

The performance of the model improved with pre-learned GCN networks (TL-F-GCN model). The curve for validation loss in Figure 7(c) for the TL F-GCN model is sloping downwards with MSE = 0.08. This value is slightly lower than the no-TL F-GCN model (validation loss MSE = 0.17). Additionally, LCE predictions on the unseen test electrolytes demonstrate significant improvement in Figure 7(d). The predicted LCE values in Figure 7(d) closely approximates actual experimental values with an MSE of 0.15, which outperforms the MSE of 0.33 reported by Kim et al. [33] using regression models on the same dataset. This result illustrates the superior scope and promise of the proposed F-GCN model in predicting properties/performance of formulations than the alternative methods based on feature selection and regression. Use of transfer learning manages to improve the accuracy of the predictive deep learning model despite being trained on small dataset. This reduced dependence on the quantity of data for deep learning is consistent with previous study by Karpov et al [71]. The predictive capability can be further enhanced by using large-scale pre-trained foundational models [69] in lieu of presently pre-trained GCN.

The major advantage of F-GCN is the ability to featurize molecules into a molecular descriptor from SMILES on the fly, thereby overcoming the need for the input feature selection process. Furthermore, by compartmentalized learning of the model with specialized smaller datasets, we induce the scope to impart selective domain knowledge to the model, thereby making the presented approach a promising solution to formulation problems in broad industry sectors. For instance, we pre-train GCN on HOMO-LUMO energy levels (HL) of electrolyte solvents and salts to encode graph representation (*GR*) (explained in Section 2) with molecular information of these quantum chemical properties through the process of back-propagation [57]. Frontier orbitals (HL) are the most critical properties of molecules that have far-reaching consequences in organic and inorganic reactivity. As per the frontier molecular orbital theory, molecules with lower LUMO levels undergo reduction reaction more easily while molecules with higher HOMO levels are oxidized first. Thus, HOMO-LUMO of electrolyte components presents an intrinsic window of the working voltage range of a battery and determines the stability of electrolytes over electrodes along with their subsequent chemical reactions [72-74]. Having such a generic impact across all battery systems, HL have been widely adopted as an important criterion to screen solvents and salts for the development of new battery electrolytes [18, 75-79]. Based on this general importance of HL for battery electrolytes, we pre-train GCN model with HL before utilizing them in the F-GCN framework for predicting the electrolyte performance. However, if the electrolyte formulation dataset targets a specialized concept such as high entropy electrolytes or high-voltage cathode stabilization [73, 80], one may use the targeted simulated property for pre-training of GCNs like $Li^+$ solvation or oxidation potential. GCN can also be trained to encode information regarding more than one molecular property. We demonstrate this concept by using another physio-chemical attribute like electric moment (EM) alongside HL as a label in GCN pre-training. EM is the measure of separation of positive and negative charges in the system, indicating the molecule's overall polarity. Recent studies have determined that the ratio of polar to non-polar constituents in electrolyte formulation has a significant say in $Li^+$ transport across the electrolyte and its subsequent desolvation phenomenon at the electrode interface [81, 82]. By pre-training GCN with more than one molecular property, we impart generalizability to the model towards different categories of electrolyte formulations, and enable explicit learning of the chemical character of a compound, thereby overcoming the lack of large molecular data during pre-training, like in the present case. Although incorporation of additional attribute knowledge may or may not improve the overall accuracy of the model. For present datasets, errors observed from TL F-GCN models pre-trained with HL alone and HL-EM were found to be similar.

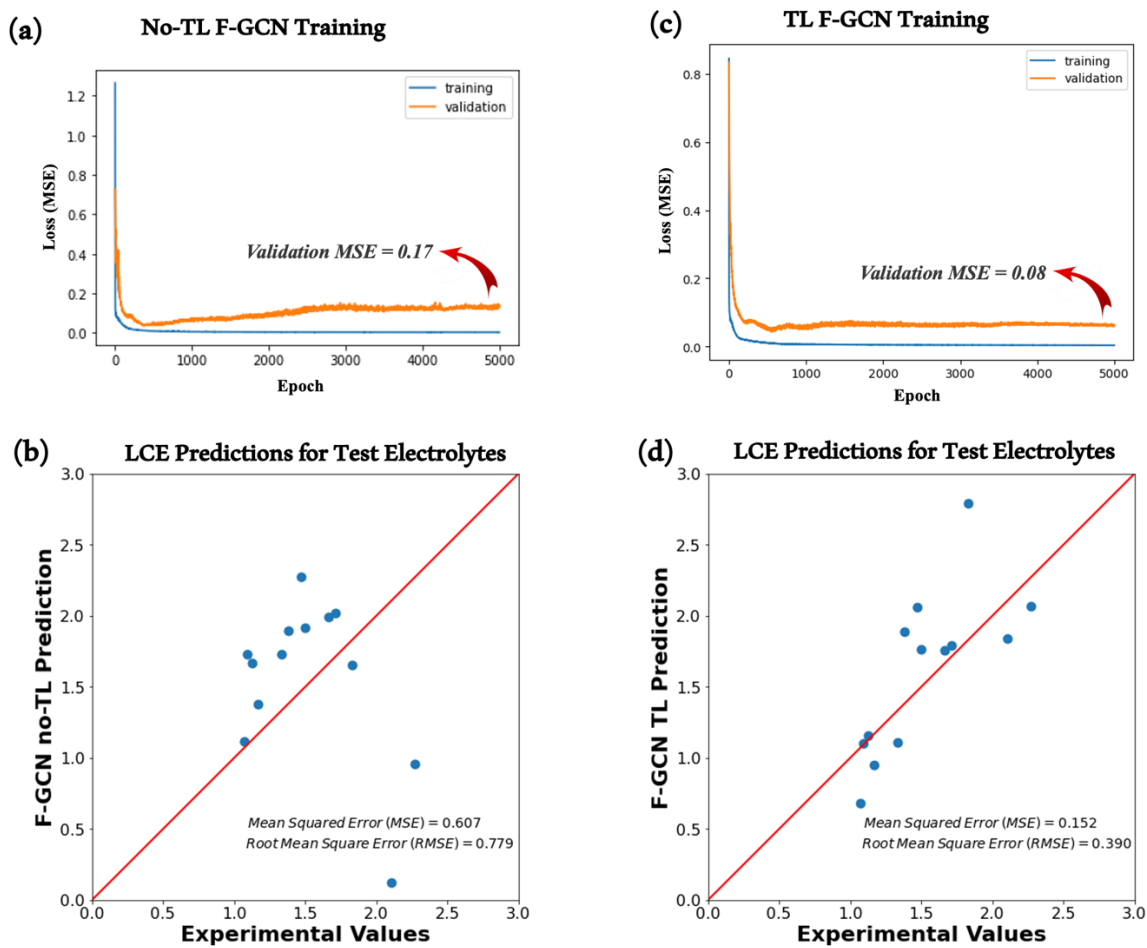

**Figure 7: (a)** Curves for loss function (MSE) calculated for training and validation data during the learning of formulations by F-GCN model (no TL) for 5000 epochs. **(b)** Parity plot showing predicted LCE values from F-GCN where no Knowledge Transfer has been implemented (no TL

F-GCN) as scatterplots with respect to the benchmark experimentally derived LCE values mapped on y=x axis. **(c)** Curves for loss function (MSE) were calculated for training and validation data during the learning of formulations by the F-GCN model using pre-trained GCN (TL F-GCN) for 5000 epochs. **(b)** Parity plot showing predicted LCE values from F-GCN using GCN pre-trained on energy levels and electric moment (TL F-GCN) as scatterplots with respect to the benchmark experimentally derived LCE values mapped on y=x axis.

*4.2. Li-I Battery Capacity Prediction*

A full-cell battery system with a functional cathode and anode has a lot more complex dependencies than a half-cell system. Currently, most data-driven approaches for battery systems are developed and tested with simplified datasets curated from different literature sources or high-throughput experimentation[33, 34, 83]. We assess the capability of the F-GCN model in predicting another important battery performance metric, i.e. specific capacity, based on variations in electrolyte formulations. The electrolyte formulation vs. specific capacity dataset was collected during our development of Oxygen-Assisted Lithium Iodine (OALI) battery, a heavy metal-free next-generation battery that promises high power and fast charging capabilities by forming a robust solid electrolyte interface (SEI)[70]. The compiled dataset effectively captures a realistic range of electrolytes for next-generation lithium-metal batteries based on the conversion chemistries.

A separate version of the F-GCN model is optimized for predicting battery capacities (in mAh/g). Hyperparameters of DNN are re-tuned for predicting the new output matrix (see Supporting Information SI-9). Optimized F-GCN has a 3-layered perceptron as external learning architecture with each layer having 1000-100-10 nodes, the last of which is connected to an output label. It took approximately 30,000 epochs for F-GCN to converge. The performance of the model is evaluated by calculating root mean squared error (RMSE) in the predicted capacities with respect to the experimentally noted capacities. It is crucial to note here that the nodal architecture of DNN became much more extensive for predicting specific capacity retained by a full battery than the simplified case of predicting CE in Li/Cu half-cell as seen in Section 4.1. This reverberates an important fact that the relationship between electrolyte and battery performance becomes much more intricate in a full battery cell. The model must take electrolyte-electrode interactions into account for accurately predicting the overall battery performance. The F-GCN model is applied to the electrolyte formulation problem with an assumption that the rest of the battery components, including electrodes, current collector, separator, and the volume of electrolyte, are mostly consistent for the dataset. The experimentally collected electrolyte formulation vs. specific capacity dataset encodes the intelligence of electrolyte-electrode interactions in the performance metric. For instance, capacity retention for electrolytes with high concentrations of polar protic solvents is consistently low as they conceptually react with electrodes deliriously. Similar trends of electrolyte component's interaction with the electrode are transmitted to the performance metric considered for learning here, with a requisite that electrode attributes are consistent for the said dataset.

Figure 8 summarizes the performance of two F-GCN models (no-TL F-GCN and TL F-GCN) trained with an electrolyte formulation dataset derived from Li-I full cell. The F-GCN model using informed molecular descriptors (TL F-GCN) outperforms the model with no knowledge transfer (no-TL F-GCN) by demonstrating relatively lower RMSE for predicted battery capacities. The predicted values from both models are plotted against experimental capacities in parity plots shown in Figure 8. The RMSE value for each plot is indicated on the figure itself. The predicted capacities in Figure 8(b) closely overlap with actual experimental values with RMSE of 20.46 mAh/g from TL F-GCN. Besides predictive precision, the model successfully segregates high-performing electrolytes from low-performing ones. The RMSE of 20-21 mAh/g in the capacity prediction is expected for the current dataset as experimental battery capacity values inherit some uncertainties based on unintended variations during battery assembly and material preparation processes. The standard deviation (cell-to-cell variation) of experimental specific capacities is in the range of 15-30 mAh/g (see SI-7) when the same electrolyte formulation is used in an OALI battery. This uncertainty in experimental values originates mainly from very slight variations observed in active material loading in cathode as indicated in section 3.3 and Supporting Information. Thus, the model predicts the performance pertaining to the electrolyte formulations with errors that are within the bounds of experimental cell-to-cell variability.

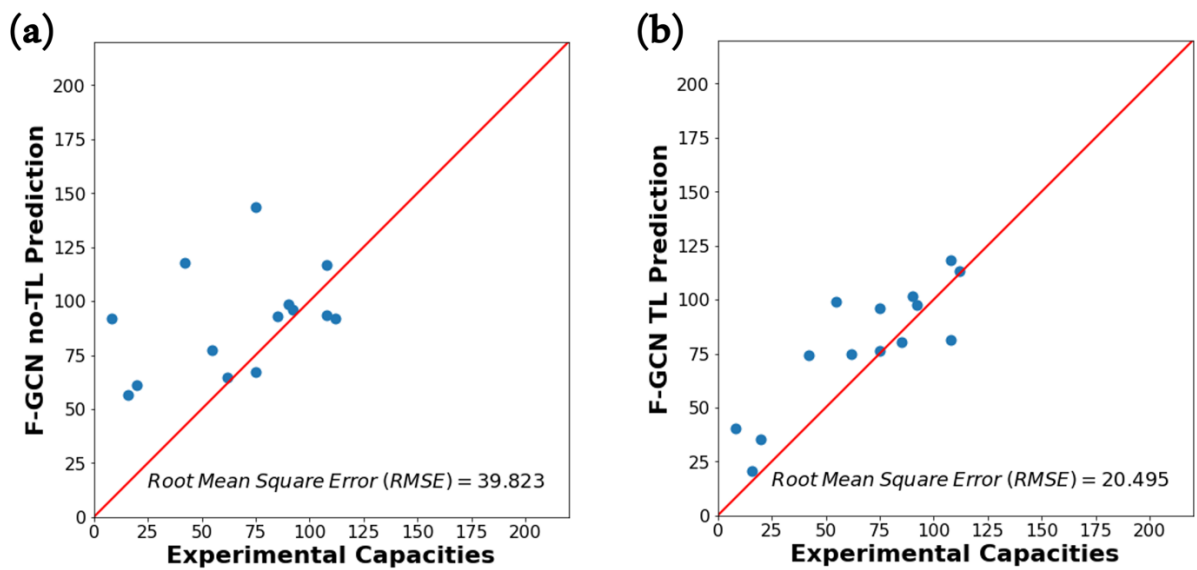

**Figure 8:** Parity plots showing predicted battery capacities (in mAh/g) as scatterplots with respect to the benchmark experimentally derived capacity values mapped on the y=x axis. **(a)** F-GCN where no Knowledge Transfer has been implemented (no-TL F-GCN), **(b)** F-GCN with HOMO-LUMO (HL) and electric moment (EM) informed molecular descriptors. RMSE values for both the predicted sets are indicated in the figures, respectively.

*4.3. Li-I Data Variance and Predictive Uncertainties*

It is evident in Figure 8(b) that the TL F-GCN model does an excellent job of recognizing high-performing and low-performing electrolyte formulations, despite oncoming uncertainties from experiments and small training data. However, notable errors are observed for low-mid capacity range (30 – 60 mAh/g) electrolyte formulations. This is indicative of biased training data that provides insufficient coverage of the full range of battery capacity, particularly, in the low-mid capacity range. The limitations of the current dataset are examined in Figure 9 which describes the distribution of the training data with the help of heat maps in terms of solvent types being used in the electrolyte formulations (Figure 9(a)) and formulant compositions (solvent and salt compositions in Figure 9(b-c)). The solvent types are depicted in the heat map with labels 0-20 and are further detailed in the color bar beside the heat map in Figure 9(a). Figure 9(d) maps the corresponding specific capacities of the battery cells. The solvent classification map in Figure 9(a) shows that the primary and the highest concentration *Solvent A* is mostly varied between 3 classes of solvents (cyclic ether, ether nitrile, and heterocyclic acetal). In contrast, significant variations in the category of co-solvent *Solvent B* are seen. The third solvent *Solvent C* is mostly used as an additive in about 12 % of the training dataset. Figure 9(b) elaborates on the molar percentage of each solvent in a respective formulation. The highest concentrations are noted for *Solvent A* (45-70 mol%), followed by *Solvent B* (5 -50 mol%) and *Solvent C* (0-20 mol%). Opposed to the solvent trends in the dataset, the 3 salts in the formulations were fixed and are lithium bis (trifluoromethyl) sulfonylimide (LiTFSI) as *Salt A*, lithium bis(oxalato)borate (LiBOB) as *Salt B*, and lithium nitrate (LiNO$_3$) as *Salt C*, respectively. *Salt C* is added to all electrolyte formulations with no significant concentration alterations (~2 mol%). Meanwhile, *Salt A* and *Salt B* are present in 40 % of the training dataset with their concentration varying from 2.5-7 mol%. Based on these limited variations in the electrolyte formulations, the battery capacities of the training data are predominantly distributed either in a higher range (> 80 mAh/g, well-performing) or close to zero (not performing) as shown in Figure 9(d). Consequently, this leads to relatively poor predictions for the low-mid capacity range. While there is acceptable disparity noted in the category of solvents in the training data as summarized in Figure 9, the scope of the model could be further improved if the training data is more inclusive of different salts and well-diversified concentration ranges. This highlights the need for an ideal design of experiments (DOE) that enables dataset sampling suitable to address a complex problem for multivariable materials space such as optimizing the formulation of a materials mixture.

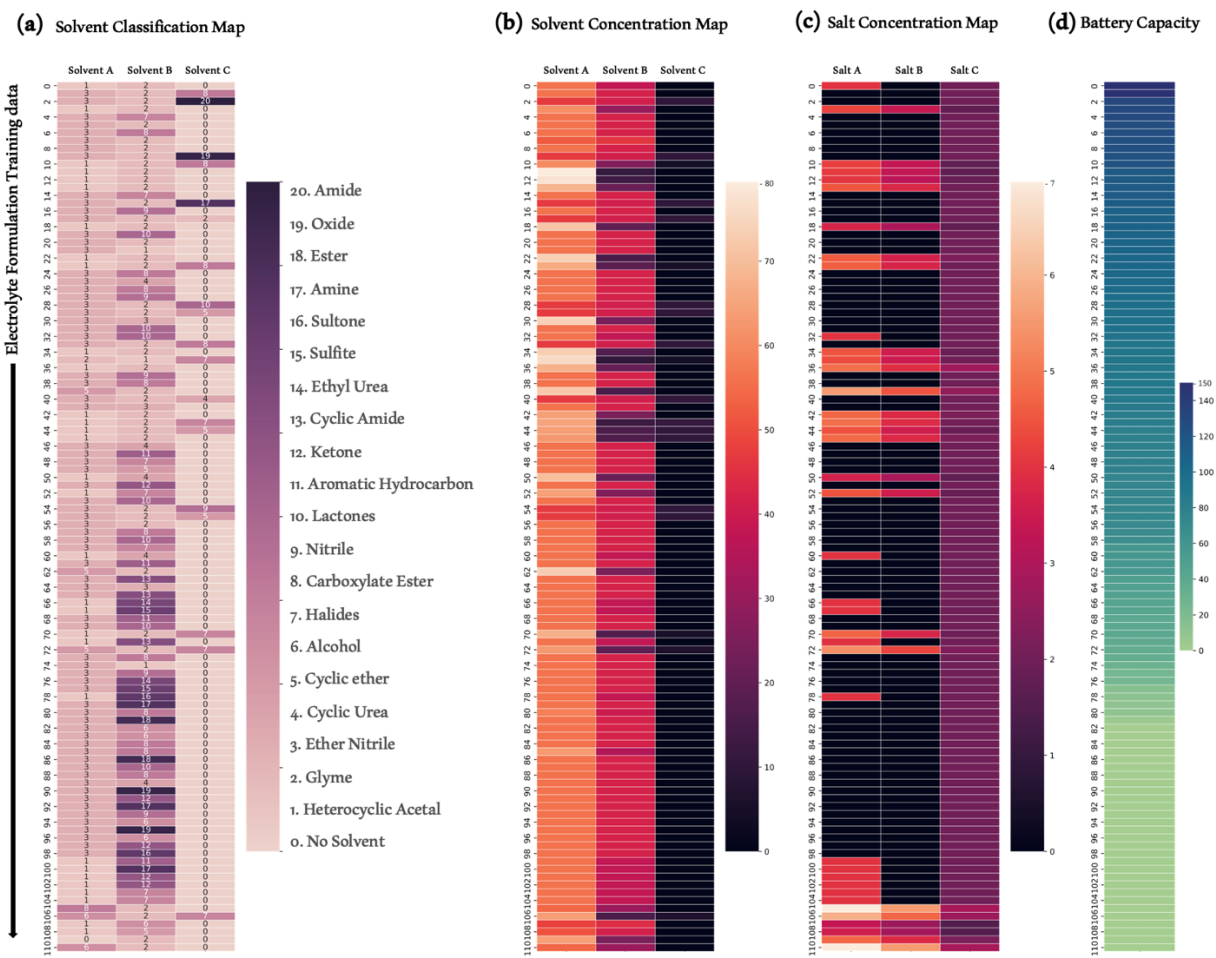

**Figure 9:** Variations in electrolyte formulation training dataset. Labels on the y-axis are identical across the plots and encode the items in the dataset. **(a)** Heat map depicting variations in the types of three electrolyte constituent solvents used in the training data. **(b)** Heat map depicting variations in molar percentage of 3 constituent solvents in the training data. **(c)** Heat map depicting variations in molar percentage of 3 constituent salts in the training data. **(d)** Heat map depicting battery capacities obtained from the experiments corresponding to the electrolyte formulations in the training data.

Lastly, we determine the uncertainty in model predictions that arises due to inherent randomness in the input observations. Being independent of the model's parameters, this uncertainty cannot be reduced by increasing the count of training data [84]. The uncertainty in the predictions from the model is evaluated by training TL F-GCN ensembles using the popular bootstrapping strategy [85, 86]. Ensembles use multiple F-GCN that have randomly initialized weights (with the exception of pre-trained GCN weights) and get trained on different bootstrap samples of the original training data. Since the original Li-I electrolyte training data (111 formulations) is too small for bootstrapping, an augmented dataset of electrolyte formulations (see SI-5 for details) is bootstrapped into random 4 sub-samples for training F-GCNs in the ensemble. We note that random initialization of DNN parameters, random shuffling of training data points, and bootstrapping are sufficient to observe predictive uncertainties. Ensembles of TL F-GCN are trained following the same procedure as described earlier in section 4.2 and are then used to predict the capacities of 14 test electrolyte formulations. The predictions from ensembles for each test sample are depicted as grey scatter points in Figure 10 and average predicted capacities are plotted by the blue line. Standard deviations in predicted capacities are summarized in Table 3. Large uncertainties are observed in the predictions for electrolytes falling in low-to-mid range capacities (30 – 60 mAh/g, test electrolyte samples 3 and 5 in Figure 10), which can be attributed to the limited coverage of such data in the training dataset as previously described. The TL F-GCN model employing the HL-EM descriptor exhibits an outstanding prediction accuracy for about 65 % of test data points (9 out of 14) with an approximate standard deviation of 10 mAh/g.

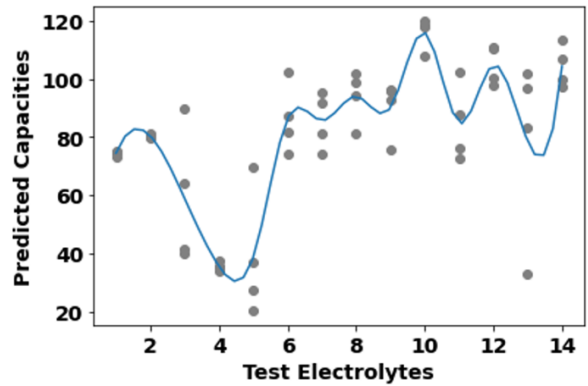

**Figure 10:** Capacities (in mAh/g) predicted by Ensemble TL F-GCNs for 14 test electrolyte formulations. Average of predicted capacities from ensembles are plotted in blue.

**Table 3** Summary of standard deviation (STD) in TL F-GCN predicted capacities from ensembles for 14 test electrolyte systems using informed descriptors, where HL-EM is HOMO-LUMO and electric moment.

| Label of Test Electrolyte Formulations | STD in F-GCN Predicted Capacities (mAh/g) |
|---|---|
| | **HL-EM** |
| 1 | 0.92 |
| 2 | 0.72 |
| 3 | 23.24 |
| 4 | 1.47 |
| 5 | 21.66 |
| 6 | 12.03 |
| 7 | 9.64 |
| 8 | 9.14 |
| 9 | 9.78 |
| 10 | 5.38 |
| 11 | 13.42 |
| 12 | 6.83 |
| 13 | 31.42 |
| 14 | 7.39 |
| **Mean STD** | **10.93** |

*4.4.   Future Scope*

With advancements in computation and data-driven approaches, there exists surface optimism about the scope of these techniques in driving the discovery and optimization of new materials. However, as we dive deeper into the solution-driven application of these techniques, there is a significant lapse in the premise and actual practice. As simulation techniques struggle with shortcomings associated with computational requirements and theoretical assumptions, data-driven methods such as machine learning are priced high due to having prerequisites of structured high-quality materials data. To address the most pressing predictive and discovery challenges among materials, a simulation-experiment-AI synergistic approach needs to be developed where an expansive simulation dataset could be streamlined with limited experimental data in order to meet the 'data' and 'knowledge' requisites of an AI model, thereby making it a more reliable in-silico solution. F-GCN aims to incorporate these requisites to solve multi-dimensional material problems as demonstrated in the case of battery electrolyte formulations. What makes this model generalizable for formulation problems across different applications is the ground concept of featurizing the structure of molecular constituents and their respective compositions into a formulation descriptor for relating to the output. This formulation descriptor could be combined with any learning algorithm to perform a variety of downstream tasks such as prediction (as demonstrated in the present work) and composition optimization. The proposed framework can be used to find the right composition of formulation constituents, especially when the design of the targeted chemical space is vast (large count of formulation constituents) and may otherwise require a brute-force experimental approach[18]. This could be obtained by affixing the formulation constituents and varying the performance label over a range of constituent compositions. A good example of this would be the integration of F-GCN in accelerated electrolyte discovery workflow where the electrolyte solvents and salts have been shortlisted with virtual high throughput screening[87] and require further compositional fine-tuning based on existing data for the battery. Due to the existence of diverse battery chemistries in the field, electrolyte formulation datasets are are mostly non-generalizable. For each set of cathode and anode, electrolyte formulations require specialized development and are highly guarded trade secrets. Therefore, a framework that can accelerate the development of electrolyte formulation design with limited experimental efforts is most welcomed in the field. As demonstrated for the OALI battery, the initial battery cell tests that are used to develop the battery chemistry could be utilized for learning and then driving the optimization of new electrolyte formulation, resulting in a fewer number of actual electrolyte optimization experiments.

Due to the underlying framework being very general, the F-GCN model can find use in applications beyond electrolytes, especially where formulation datasets are scarce and we need to incorporate domain knowledge to enhance accuracy. By pretraining molecular structures on labels identified as critical in the subject matter, we couple learning molecular geometries with domain knowledge transfer. Hence, the resulting molecular descriptors could be made unique for the problem system. We demonstrate the proof of concept by pre-training the molecular structure model with a small, specialized simulation dataset that contained targeted electrolyte molecules. The accuracy of the formulation model could be further improved by describing molecules with pre-trained foundational models such as MoLFormer [69] in substitute of GCN.

## 5. Conclusion

In conclusion, we propose a graph-based deep learning model, F-GCN, which maps structure-composition-performance relationships within the formulation space. The F-GCN model is constituted of six graphs for formulation constituents and an external learning architecture. The model is trained within two domains: the learning of molecular graphs and the learning of formulations. The proposed approach is tested with two different electrolyte formulation vs. performance datasets: Li/Cu half-cell data sourced from the literature, and Li-I full-cell data experimentally acquired in our lab. The TL F-CGN model trained and validated with the Li/Cu half-cell data, exhibited enhanced accuracy in predicting logarithmic Coulombic efficiency with MSE of 0.15, surpassing the performance of previous ML approaches in the literature including regression and kernel methods [33]. The TL F-CGN model also demonstrated outstanding predictive performance for battery capacities of Li-I full cells, achieving an RMSE of 20 mAh/g and an approximated standard deviation of 10 mAh/g within the ensembles. Given the inherent cell-to-cell variations resulting from experimental cell assembly and operation of these conversion batteries, the accuracy of the F-GCN prediction model depends more on the quality of the training data rather than the quantity. The proposed model attempts to simplify data-driven discovery and optimization of mixed materials such as formulations, where data poses unstructured relationships.

**SUPPORTING INFORMATION**

Molecular Graph Representation; Graph convolution networks (GCN); F-GCN; Li/Cu half- cell electrolyte dataset; Li-I full cell dataset; Battery experiment details; Cell-to-cell variability in experiments; Hyperparameter tuning for LCE prediction; Hyperparameter tuning for capacity prediction.

## AUTHOR CONTRIBUTIONS

V.S., M.G., Y.H.L and D.Z. devised the idea of the project. V.S. implemented the idea and performed primary coding, model training, validation, and manuscript preparation. M.G., A.T., K.N. and L.S. performed the battery experiments to collect the training and testing data for the study. V.S. collected the simulation data using open-sourced DFT simulation workflows on Simulation Toolkit for Scientific Discovery (ST4SD). D.C curated data from literature sources for benchmarking.

## CONFLICT OF INTEREST STATEMENT

The authors have no conflicts of interest to declare. All authors have seen and agree with the contents of the manuscript and there is no financial interest to report. We certify that the submission is original work.

## CODE AVAILABILITY

Code used in the study is presented in Supporting Information. The simulation workflows and tools by Simulation Toolkit for Scientific Discovery (ST4SD) are used to generate simulation data in this study which are now available to open-source community at https://st4sd.github.io/overview/. Restrictions may apply to the availability of the simulation codes which are protected under the software license.

## DATA AVAILABILITY

The experimental data and simulation data used in this paper can be found along with the Supporting Information.